\documentclass[
prd 
,preprintnumbers%
,showpacs ,amssymb,superscriptaddress,aps
]{revtex4}
\usepackage{graphicx}
\input{epsf}

\usepackage{amsmath,amssymb}
\usepackage{bm}

\newcommand{\dalm}{\kern1pt\vbox{\hrule height 0.9pt\hbox{\vrule width 0.9pt
\hskip 2.5pt\vbox{\vskip 5.5pt}\hskip 3pt\vrule width 0.3pt}\hrule height 0.3pt}
\kern1pt}

\begin{document}
\preprint{YITP-14-50}



\title{Stellar oscillations in Eddington-inspired Born-Infeld gravity}

\author{Hajime Sotani}
\email{sotani@yukawa.kyoto-u.ac.jp}
\affiliation{Yukawa Institute for Theoretical Physics, Kyoto University, Kyoto 606-8502, Japan
}

\date{\today}

\begin{abstract}
We consider the stellar oscillations of relativistic stars in the Eddington-inspired Born-Infeld gravity (EiBI). In order to examine the specific frequencies, we derive the perturbation equations governing the stellar oscillations in EiBI by linearizing the field equations, and numerically determine the oscillation frequencies as changing the coupling parameter in EiBI, $\kappa$, and stellar models. As a result, we find that the frequencies depend strongly on the value of $\kappa$, where the frequencies in EiBI with negative $\kappa$ become higher and those with positive $\kappa$ become lower than the expectations in general relativity. We also find that, via the observation of the fundamental frequency, one could distinguish EiBI with $8\pi\epsilon_0|\kappa|\gtrsim 0.03$ from general relativity, independently of the equation of state (EOS) for neutron star matter, where $\epsilon_0$ denotes the nuclear saturation density and $\epsilon_0\kappa$ become dimensionless parameter. With the further constraints on EOS, one might distinguish EiBI even with $8\pi\epsilon_0|\kappa|\lesssim 0.03$ from general relativity.
\end{abstract}

%
\maketitle
\section{Introduction}
\label{sec:I}

Asteroseismology is a unique approach to investigate stellar properties, which is similar to helioseismology for the Sun. This is a technique to see the stellar properties by using the observable information of stellar oscillations. Via the observations of spectra of oscillation frequencies, one expects to find the stellar mass, radius, equation of state (EOS), spin frequency, and information about magnetic field (e.g., \cite{AK1996, AK1998,STM2001,SH2003,PA2011,SYMT2011,DGKK2013}). In practice, the possibilities to constrain the saturation parameters of nuclear matter are also suggested, using the quasi-periodic oscillations observed in the giant flare phenomena \cite{SW2009,GNJL2011,S2011,SNIO2012}, whose sources are considered as strongly magnetized neutron stars \cite{DT1992}. Furthermore, the direct observations of gravitational waves induced by the stellar oscillations might enable us to probe the gravitational theory in strong-field regime \cite{SK2004,S2009,YYT2012,S2014a}. Many experiments and observations in weak-field regime, such as the solar system, tell us the validity of general relativity, while the tests of gravitational theory in strong-field regime are still poor. That is, the gravitational theory in strong-field regime might be different from general relativity. If so, one could probe the gravitational theory through the observations associated with compact objects \cite{W1993,P2008}. Verification of gravitational theory is one of the importances to directly detect the gravitational waves.

Eddington-inspired Born-Infeld gravity (EiBI) \cite{EiBI} recently attracts attention as a modified gravitational theory, because the big bang singularity can be avoided with this theory. EiBI is based on the gravitational action proposed by Eddington \cite{E1924} and on the nonlinear electrodynamics by Born and Infeld \cite{BI}. EiBI becomes completely equivalent to general relativity in vacuum, while EiBI can deviate from general relativity in the presence of matter. Because the gravity in EiBI is nonlinearly coupled with matter, one can expect the significant deviation in the high density region, such as inside the compact objects. Actually, the spherically symmetric neutron star models in EiBI have been constructed, which can deviate from the predictions in general relativity even for the low-mass neutron stars \cite{PCD2011,PDC2012,SLL2012,SLL2013,HLMS2013,S2014}. That is, via the direct measurements of stellar mass and radius, one might be able to distinguish EiBI from general relativity.

On the other hand, as mentioned before, the frequencies of compact objects could tell us the information associated with the compact objects. If the spectra of stellar oscillations expected in EiBI would become different from those in general relativity, one might be possible to distinguish the gravitational theory via the observation of stellar oscillations such as the gravitational waves radiated from the compact objects. So, in this paper, we consider the stellar oscillations in EiBI. In particular, to examine the oscillation frequencies, we adopt the relativistic Cowling approximation, where the metric is assumed to be fixed during the oscillations. Then, as changing the coupling parameter in EiBI and stellar models, we will examine the spectra systematically. This paper is organized as follows. In the next section, we briefly summarized EiBI and equilibrium stellar models in EiBI. In Sec. \ref{sec:III}, we derive the perturbation equations describing the stellar oscillations and solve it numerically. Finally, we make a discussion in Sec. \ref{sec:IV}. In this paper, we adopt geometric units, $c=G=1$, where $c$ and $G$ denote the speed of light and the gravitational constant, respectively, and the metric signature is $(-,+,+,+)$.

\section{Stellar Equilibrium in EiBI}
\label{sec:II}

In this section, we briefly mention EiBI and the relativistic stellar models in EiBI, where we especially consider the spherically symmetric stellar models. EiBI is proposed by Ba\~nados and Ferreira \cite{EiBI}, which can be obtained with the action as
\begin{equation}
   S = \frac{1}{16\pi}\frac{2}{\kappa} \int d^4x \left(\sqrt{|g_{\mu\nu} + \kappa R_{\mu\nu}|} - \lambda\sqrt{-g}\right)
     + S_{\rm M}[g,\Psi_{\rm M}],
\end{equation}
where $|g_{\mu\nu} + \kappa R_{\mu\nu}|$ and $g$ denote the determinants of $(g_{\mu\nu} + \kappa R_{\mu\nu})$ and $g_{\mu\nu}$, while $R_{\mu\nu}$ is the Ricci tensor constructed with the connection $\Gamma^\mu_{\alpha\beta}$. We remark again that the connection $\Gamma^{\mu}_{\alpha\beta}$ should be considered as the independent field from the metric tensor $g_{\mu\nu}$ in EiBI. The matter action $S_{\rm M}$ depends on the metric and matter field $\Psi_{\rm M}$. This theory has two parameter $\lambda$ and $\kappa$. The dimensionless constant $\lambda$ is associated with the cosmological constant $\Lambda$, such as $\lambda = 1 + \kappa \Lambda$. In this paper, we consider only asymptotically flat solutions, i.e., we adopt that $\lambda=1$. The remaining parameter $\kappa$ is the Eddington parameter, which is constrained in the context of the observations in solar system, big bang nucleosynthesis, and the existence of neutron stars \cite{EiBI,kappa01,kappa02,PCD2011}. Additionally, terrestrial measurements of the neutron skin thickness of ${}^{208}$Pb and astronomical observations of the radius of $0.5M_\odot$ neutron star could enable us to constrain $\kappa$ \cite{S2014}.

The field equations are obtained by varying the action \cite{EiBI};
\begin{gather}
  \Gamma^\mu_{\alpha\beta} = \frac{1}{2}q^{\mu\sigma}\left(q_{\sigma\alpha,\beta} + q_{\sigma\beta,\alpha}
     - q_{\alpha\beta,\sigma}\right), \label{eq:1} \\
  q_{\mu\nu} = g_{\mu\nu} + \kappa R_{\mu\nu}, \label{eq:2} \\
  \sqrt{-q}q^{\mu\nu} = \sqrt{-g}g^{\mu\nu} - 8\pi\kappa\sqrt{-g}T^{\mu\nu}, \label{eq:3}
\end{gather}
where $q_{\mu\nu}$ and $q$ denote an auxiliary metric associated with the physical metric $g_{\mu\nu}$ via Eq. (\ref{eq:2}) and its determinant, while $T^{\mu\nu}$ is energy-momentum tensor defined with the matter action $S_{\rm M}$ as
\begin{equation}
  T^{\mu\nu} = \frac{1}{\sqrt{-g}}\frac{\delta S_{\rm M}}{\delta g_{\mu\nu}}.
\end{equation}
With the covariant derivative $\nabla_\mu$, which is defined with $g_{\mu\nu}$, the energy-momentum conservation law is expressed as $\nabla_\mu T^{\mu\nu}=0$. From Eq. (\ref{eq:3}), one can show that the physical metric $g_{\mu\nu}$ is completely equivalent to the auxiliary metric $q_{\mu\nu}$, when $T^{\mu\nu}=0$.

The structures of neutron stars in EiBI have been discussed in some literatures \cite{PCD2011,PDC2012,SLL2012,SLL2013,HLMS2013,S2014}. The metric for the spherically symmetric objects is expressed as
\begin{gather}
  g_{\mu\nu}dx^\mu dx^\nu = -e^{\nu}dt^2 + e^{\lambda}dr^2 + f(d\theta^2 + \sin^2\theta d\phi^2), \\
  q_{\mu\nu}dx^\mu dx^\nu = -e^{\beta}dt^2 + e^{\alpha}dr^2 + r^2(d\theta^2 + \sin^2\theta d\phi^2),
\end{gather}
where $\nu$, $\lambda$, $\beta$, $\alpha$, and $f$ are functions of $r$. Assuming that the neutron stars are composed of perfect fluid, the energy-momentum tensor is given by
\begin{equation}
  T^{\mu\nu} = \left(\epsilon + p\right)u^\mu u^\nu + p g^{\mu\nu},
\end{equation}
where $\epsilon$ and $p$ are the energy density and pressure, while $u^{\mu}$ corresponds to the four velocity of matter given as $u^{\mu}=(e^{-\nu/2},0,0,0)$. Then, from Eqs. (\ref{eq:2}), (\ref{eq:3}), and the energy-momentum conservation law, one can obtain the Tolman-Oppenheimer-Volkoff (TOV) equations in EiBI \cite{PCD2011,PDC2012,SLL2012,SLL2013,HLMS2013,S2014}. To close the equation system, one needs prepare the relationship between the pressure and density, i.e., EOS. In particular, in this paper, we adopt two realistic EOSs to construct the neutron star models, i.e, Shen EOS \cite{ShenEOS} and FPS EOS \cite{FPS}. Shen EOS is based on the relativistic mean field approach, while FPS EOS is based on the Skyrme-type effective interaction (see \cite{SIOO2014} for more details about the adopted EOSs). Note that the appearance of the curvature instabilities at the stellar surface constructed with a polytropic EOS is pointed out \cite{PS2012}, which could be a problem to solve. Furthermore, the coupling constant $\kappa$ is constrained from the evidence that compact objects exist \cite{PCD2011}, i.e.,
\begin{gather}
  8\pi p_c\kappa < 1\ \  {\rm for}\ \  \kappa>0, \\
  8\pi\epsilon_c |\kappa| <1\ \  {\rm for} \ \ \kappa<0,
\end{gather}
where $p_c$ and $\epsilon_c$ denote the central pressure and density. Hereafter, we adopt $8\pi\epsilon_0\kappa$ as a normalized coupling constant, where $\epsilon_0$ is the nuclear saturation density given by $2.68\times 10^{14}$ g cm$^{-3}$. We remark that the coupling constant $\kappa$ has been constrained from the observations in the solar system, i.e., $|\kappa| \lesssim 3\times 10^5$ m$^5$ s$^{-2}$ kg$^{-1}$ \cite{kappa01}, which leads to $|8\pi\epsilon_0\kappa|\lesssim 2.25\times 10^7$.


\begin{figure}
\begin{center}
\includegraphics[scale=0.53]{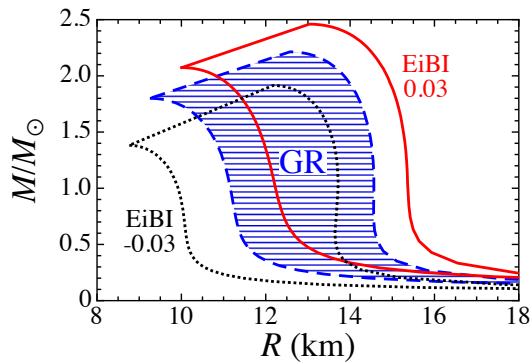} 
\end{center}
\caption{
(Color online) Comparison between the neutron star mass-radius relations in general relativity and in EiBI with $8\pi\epsilon_0\kappa=\pm0.03$. The shaded region surrounded by the broken line shows the allowed values of mass and radius for EOS with stiffness between FPS and Shen EOSs in general relativity, while the regions surrounded by the solid and dotted lines show those in EiBI with $8\pi\epsilon_0\kappa=0.03$ and $-0.03$.
}
\label{fig:MR}
\end{figure}

In Fig. \ref{fig:MR}, we show the mass-radius relations in general relativity and in EiBI with $8\pi\epsilon_0\kappa=\pm0.03$, where the shaded region surrounded by the broken line corresponds to the allowed region in mass-radius relation for EOS with stiffness between FPS and Shen EOSs, while the regions surrounded by the solid and dotted lines correspond to those in EiBI with $8\pi\epsilon_0\kappa=0.03$ and $-0.03$.
From this figure, one can observe a large uncertainty in the mass-radius relation due to EOS, compared with that due to the gravitational theory. In practice, even if $8\pi\epsilon_0|\kappa|\simeq 0.03$, it might be difficult to distinguish the gravitational theory by using the measurements of stellar mass and radius.

\section{Spectra of Stellar Oscillations}
\label{sec:III}

In this paper, as mentioned before, we focus on the stellar oscillations of the relativistic stars in EiBI. For this purpose, we adopt the Cowling approximation as a first step, i.e., we neglect the metric perturbations as $\delta q_{\mu\nu}=\delta g_{\mu\nu}=0$. The Lagrangian displacement vector of matter element $\xi^i$ is given by
\begin{eqnarray}
  \xi^i &=& \left(\xi^r,\xi^\theta,\xi^\phi\right) \nonumber \\
           &=&  \left(W,-V\partial_\theta,-V\sin^{-2}\theta\partial_\phi\right)\frac{1}{r^2}Y_{\ell m},
\end{eqnarray}
where $W$ and $V$ correspond to functions of $t$ and $r$, while $Y_{\ell m}$ denotes the spherical harmonics. With such variables, the perturbation of four-velocity $\delta u^{\mu}$ can be expressed as
\begin{equation}
  \delta u^{\mu} = \left(\dot{W},-\dot{V}\partial_\theta,-\dot{V}\sin^{-2}\theta\partial_\phi\right)\frac{1}{r^2}e^{-\nu/2}Y_{\ell m},
\end{equation}
where the dot denotes partial derivative with respect to $t$. Additionally, the perturbations of energy density and pressure are given by
\begin{equation}
 \delta\epsilon = \delta \epsilon(t,r)Y_{\ell m}\ \ {\rm and} \ \ \delta p = \delta p(t,r)Y_{\ell,m}.
\end{equation}

Then, the perturbation equations in the Cowling approximation can be derived from the variation of the energy-momentum conservation law, i.e., $\nabla_\mu(\delta T^{\mu\nu})=0$. In practice, one can obtain the following equations;
\begin{gather}
  r^2\delta\epsilon + \epsilon' W + (\epsilon+p)\left[W' + \left(\frac{\lambda'}{2}+\frac{f'}{f}-\frac{2}{r}\right)W
     + \ell(\ell+1)V\right] = 0, \label{eq:a} \\
  \frac{\epsilon+p}{r^2}e^{-\nu} \ddot{W} + e^{-\lambda}\delta p' + \frac{\nu'}{2}e^{-\lambda}\left(\delta p + \delta\epsilon\right) = 0,
      \label{eq:b} \\
  -\frac{\epsilon + p}{r^2}e^{-\nu}\ddot{V} + \frac{1}{f}\delta p = 0, \label{eq:c}
\end{gather}
where the prime denotes partial derivative with respect to $r$. In addition to the above equations, one can show that $\delta p$ is associated with $\delta \epsilon$ as $\delta p=c_s^2\delta\epsilon$, where $c_s$ denotes the sound speed. At last, combining Eqs. (\ref{eq:a}) -- (\ref{eq:c}) with the relation of $\delta p=c_s^2\delta\epsilon$, one can get the perturbation equations for $W$ and $V$ as
\begin{gather}
  W' = \left[\frac{\nu'}{2c_s^2} - \frac{\lambda'}{2} - \frac{f'}{f} + \frac{2}{r}\right]W
      + \left[\frac{\omega^2}{c_s^2} fe^{-\nu} -  \ell(\ell+1)\right]V, \label{eq:A1} \\
  V' = -\frac{1}{f}e^{\lambda}W + \left[\frac{2}{r} - \frac{f'}{f} + \nu'\right]V.  \label{eq:A2}
\end{gather}
where we assume that the perturbation variables have a harmonic time dependence, such as $W(t,r)=W(r)e^{i\omega t}$.

With the appropriate boundary conditions, the problem to solve becomes the eigenvalue problem with respect to $\omega$. The boundary condition at the stellar surface is that the Lagrangian perturbation of pressure should be vanished, i.e., $\Delta p=0$, which reduces to
\begin{equation}
  2f\omega^2 e^{-\nu}V + \nu'W = 0.
\end{equation}
On the other hand, the perturbation variables should be regular at the stellar center. Using Eqs. (\ref{eq:A1}) and (\ref{eq:A2}), one can show that $W$ and $V$ should behave in the vicinity of stellar center as
\begin{equation}
  W = Cr^{\ell+1}\ \ {\rm and}\ \ V=-Cr^{\ell}/\ell,
\end{equation}
where $C$ is a constant. Hereafter, we especially focus on the $\ell=2$ modes, which can be dominating modes in gravitational wave radiations from the compact objects.

\begin{figure*}
\begin{center}
\begin{tabular}{cc}
\includegraphics[scale=0.5]{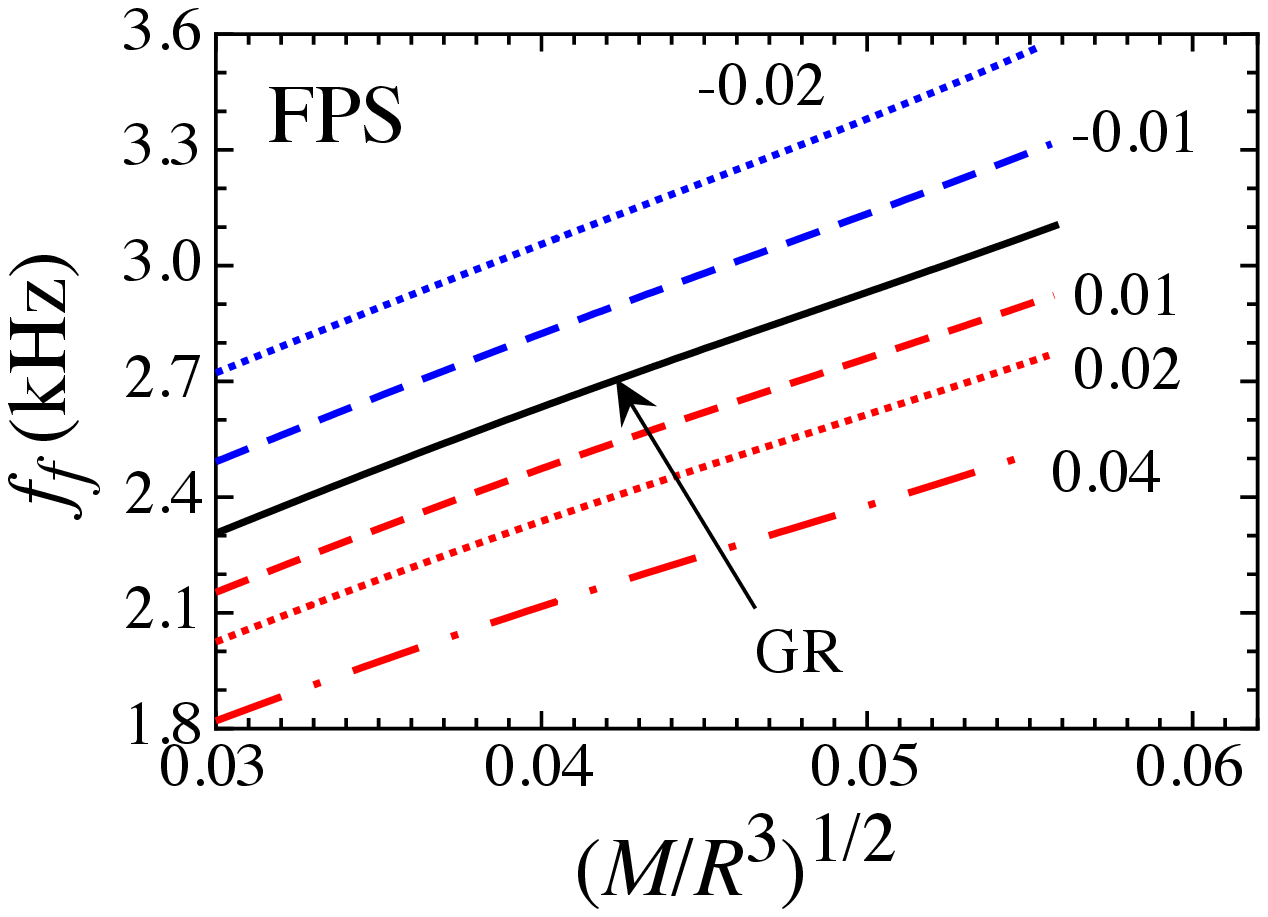} &
\includegraphics[scale=0.5]{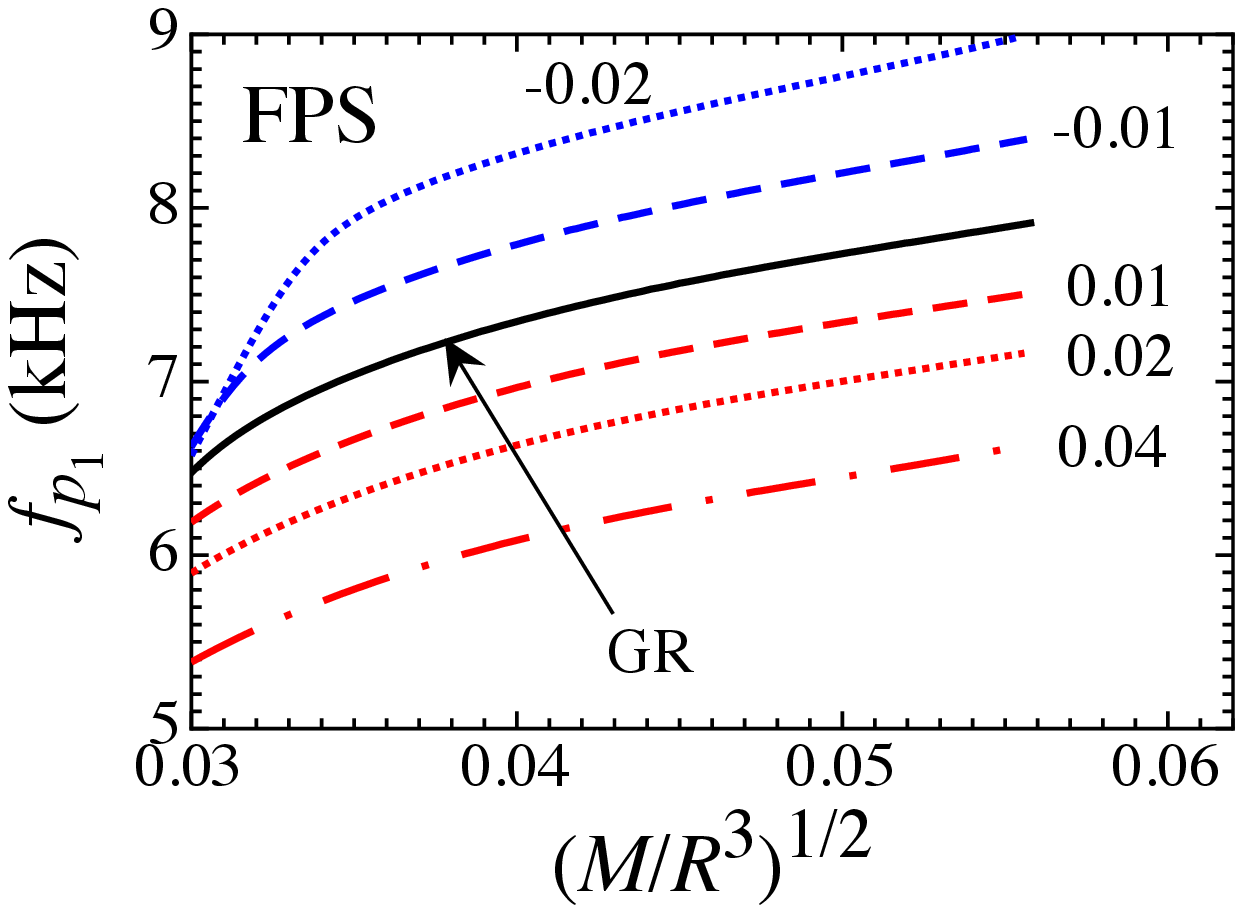}
\end{tabular}
\end{center}
\caption{
(Color online) With FPS EOS, the frequencies of $f$ mode (left panel) and $p_1$ mode (right panel) are shown as a function of the stellar average density $(M/R^3)^{1/2}$. In both panels, the frequencies in general relativity ($\kappa=0$), while the broken, dotted, and dot-dashed lines are corresponding to the results in EiBI with $8\pi\epsilon_0|\kappa|=0.01$, 0.02, and $8\pi\epsilon_0\kappa=0.04$.
}
\label{fig:FPS}
\end{figure*}

First, in order to see the dependence of the oscillation frequencies in EiBI with different values of $\kappa$, we calculate it with a specific EOS, i.e., FPS EOS. Fig. \ref{fig:FPS} shows the $f$ mode frequencies in the left panel and the $p_1$ mode frequencies in the right panel as a function of the stellar average density $(M/R^3)^{1/2}$, where the frequencies are calculated with FPS EOS. We remark that $(M/R^3)^{1/2}=3.46\times 10^{-2}$ km$^{-1}$ for a typical stellar model with $R=12$ km and $M=1.4M_\odot$. In this figure, the solid line corresponds to the frequencies in general relativity ($\kappa=0$), while the broken, dotted, and dot-dashed lines are corresponding to the results in EiBI with $8\pi\epsilon_0|\kappa|=0.01$, 0.02, and $8\pi\epsilon_0\kappa=0.04$, respectively. In the both modes, one can see  that the frequencies with negative $\kappa$ deviate more from the results in general relativity, compared with the frequencies with positive $\kappa$. In practice, for the typical stellar model with $(M/R^3)^{1/2}=0.0346$, the frequencies of $f$ mode in EiBI with $8\pi\epsilon_0\kappa=-0.01$ and $-0.02$ become 7.5\% and 16.8\% larger than that in general relativity, while those in EiBI with $8\pi\epsilon_0\kappa=0.01$, $0.02$, and 0.04 become 6.3\%, 11.6\%, and 20.4\% smaller than that in general relativity. Also, the frequencies of $p_1$ mode in EiBI with $8\pi\epsilon_0\kappa=-0.01$ and $-0.02$ become 6.0\% and 12.4\% larger than that in general relativity, while those in EiBI with $8\pi\epsilon_0\kappa=0.01$, $0.02$, and 0.04 become 5.3\%, 9.8\%, and 17.7\% smaller than that in general relativity. Additionally, we emphasize that the deviation of frequencies from the predictions in general relativity could depend on the gravitational theory, although the frequencies in EiBI may partially degenerate to those in another gravitational theory (cf., the results in scalar tensor gravity \cite{SK2004}). Thus, one may be able to distinguish EiBI from scalar-tensor gravity by collecting the observational data radiated from several neutron stars, if the observed frequencies would deviate from the predictions in general relativity.

\begin{figure*}
\begin{center}
\begin{tabular}{cc}
\includegraphics[scale=0.5]{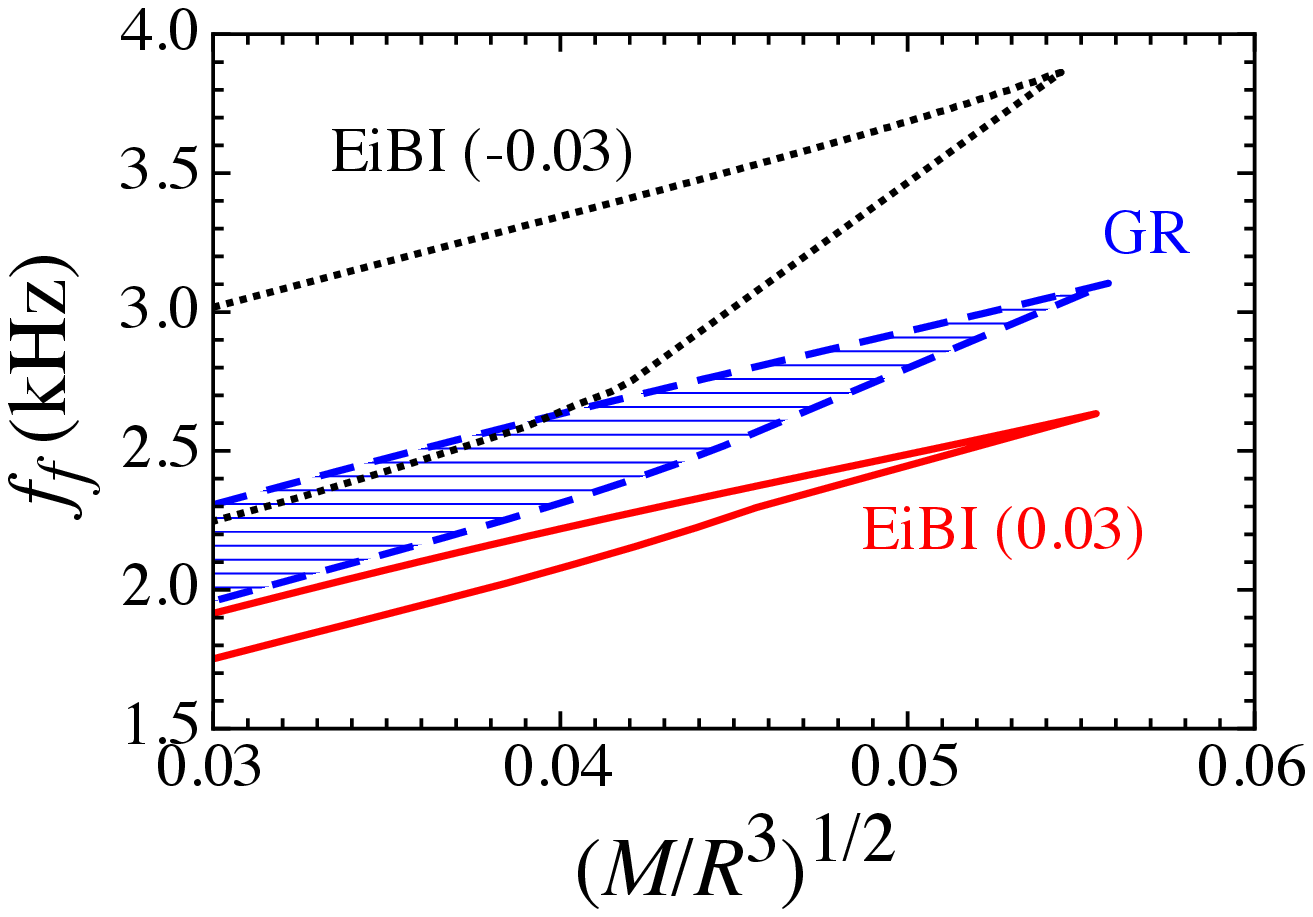} &
\includegraphics[scale=0.5]{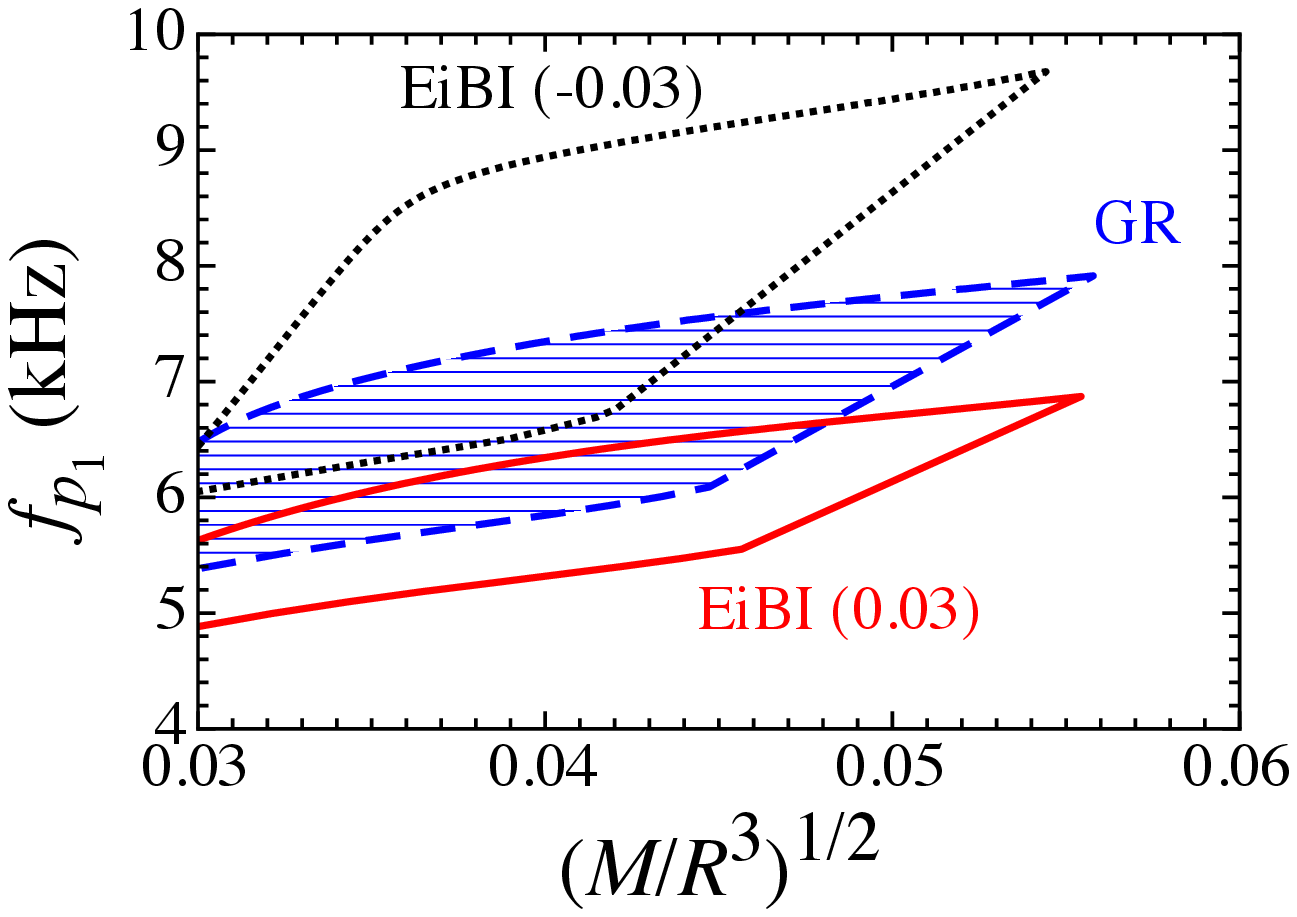}
\end{tabular}
\end{center}
\caption{
(Color online) Uncertainties due to the adopted EOS in the $f$ mode (left panel) and $p_1$ mode (right panel) frequencies, where the shaded region surrounded by the broken line denotes the expected frequencies in general relativity, while the regions surrounded by the solid and dotted lines correspond to those in EiBI with $8\pi\epsilon_0\kappa=0.03$ and $-0.03$.
}
\label{fig:l2}
\end{figure*}

From the observational point of view, as shown in Fig. \ref{fig:MR}, one might have to take into account the uncertainty due to EOS. In Fig. \ref{fig:l2}, we show the $f$ mode frequencies (left panel) and $p_1$ mode frequencies (right panel) both in general relativity and in EiBI with $8\pi\epsilon_0\kappa=\pm0.03$ as a function of the stellar average density. In the both panels, the shaded regions surrounded by the broken lines denote the frequencies expected for EOS with stiffness between FPS and Shen EOSs in general relativity, while the regions surrounded by the solid and dotted lines denote those in EiBI with $8\pi\epsilon_0\kappa=0.03$ and $-0.03$. Comparing to the mass-radius relation shown in Fig. \ref{fig:MR}, one can observe that the frequencies depend weakly on the EOS. This could be because that the $f$ mode oscillation, which is an acoustic wave, propagates inside the star with sound velocity associated with the stellar average density. In fact, it has been suggested in general relativity that the $f$ mode frequencies are written as a linear function of the stellar average density, which weakly depends on the adopted EOS \cite{AK1996,AK1998}. From the left panel in Fig. \ref{fig:l2}, one can obviously see that the $f$ mode frequencies in EiBI with $8\pi\epsilon_0\kappa\simeq\pm0.03$  could be distinguished from those in general relativity, even if the uncertainty in frequencies due to EOS would exist. That is, via the direct observations of $f$ mode oscillations, one could distinguish EiBI from general relativity, if $8\pi\epsilon_0|\kappa|\gtrsim 0.03$, independently of EOS for neutron star matter. Of course, if the EOS for neutron star matter would be determined or constrained via the other astronomical observations and/or terrestrial unclear experiments, one might distinguish EiBI even with $8\pi\epsilon_0|\kappa|\lesssim 0.03$ from general relativity. On the other hand, with the uncertainty due to EOS, it seems to be difficult to distinguish EiBI with $8\pi\epsilon_0\kappa\simeq 0.03$ from general relativity via the observations of $p_1$ mode oscillations.

\section{Conclusion}
\label{sec:IV}

Eddington-inspired Born-Infeld gravity (EiBI) attracts attention as a modified gravitational theory in the context of avoiding the big bang singularity. This theory completely agrees with general relativity in vacuum, but can deviate from general relativity in the region with matter. In this paper, we especially forces on the stellar oscillations in EiBI, and examine the oscillation frequencies of neutron stars as changing the Eddington parameter $\kappa$. For this purpose, we derive the perturbation equations with relativistic Cowling approximation by linearizing the energy-momentum conservation law. As a result, we find that the $f$ and $p_1$ mode frequencies depend strongly on the Eddington parameter. Compared with the expectations in general relativity ($\kappa=0$), the frequencies in EiBI with negative $\kappa$ become high and those with positive $\kappa$ become low. Additionally, in general, there exists an uncertainty in stellar models due to EOS of neutron star matter, but we show that one could identify EiBI with $8\pi\epsilon_0|\kappa|\gtrsim 0.03$ from general relativity independently of the adopted EOS. Furthermore, one might be able to distinguish EiBI even with $8\pi\epsilon_0|\kappa|\lesssim 0.03$ from general relativity, if the EOS would be constrained from the astronomical observations and/or terrestrial nuclear experiments. In this paper, although we adopt the relativistic Cowling approximation as a first step, we will do more complex analysis for gravitational waves radiated from neutron stars in EiBI without such approximation somewhere. In fact, the damping time of gravitational waves is also one of the important information from the asteroseismological point of view. Such an additional information must help us to constrain the gravitational theory more clearly.

\acknowledgments
This work was supported 
by Grants-in-Aid for Scientific Research on Innovative Areas through No.\ 24105001 and No.\ 24105008 provided by MEXT, 
by Grant-in-Aid for Young Scientists (B) through No.\ 24740177 and No.\ 26800133 provided by JSPS,
by the Yukawa International Program for Quark-hadron Sciences, and 
by the Grant-in-Aid for the global COE program ``The Next Generation of Physics, Spun from Universality and Emergence" from MEXT.



\end{document}